\documentclass[twocolumn,aps,superscriptaddress,nolongbibliography]{revtex4-2}
\usepackage[utf8]{inputenc}
\usepackage{graphicx}
\usepackage{amsmath}
\usepackage{amssymb}
\usepackage{bm}
\usepackage{multirow} 
\usepackage{booktabs}
\usepackage{hyperref}

\begin{document}

\title{Rotational-state dependence of interactions between polar molecules}
\author{Etienne F. Walraven}
\affiliation{Institute for Molecules and Materials, Radboud University, Nijmegen, The Netherlands}
\author{Tijs Karman}
\email{tkarman@science.ru.nl}
\affiliation{Institute for Molecules and Materials, Radboud University, Nijmegen, The Netherlands}

\begin{abstract}
   The long-range electrostatic interactions between molecules depend strongly on their relative orientation, which manifests as a rotational state dependence.
   Interactions between molecules in the same rotational quantum state are well-known attractive rotational van der Waals interactions.
   Interactions in rotational states that differ by one quantum show resonant dipole-dipole interactions. 
   We show that where molecules are in rotational states that differ by more than one quantum, they exhibit repulsive van der Waals interactions. 
   At temperatures below a millikelvin, this effect can reduce collisional loss by multiple orders of magnitude.
   These repulsive interactions lead to applications in quantum simulation and impurity physics with ultracold polar molecules.
\end{abstract}

\maketitle

\section{Introduction \label{sec:intro}}

Well-separated molecules interact with one another through the Coulomb potential set by their electric fields. This electric field can be described as an expansion in terms of multipoles, such as the molecular charge, dipole moment, quadrupole moment, and so on \cite{stone:96,gray:76}. The electrostatic interaction between two molecules can then be expanded similarly, giving rise to dipole-dipole, dipole-quadrupole, quadrupole-quadrupole interactions, and so forth. Interactions involving higher multipole moments decay with higher inverse powers of the intermolecular distance $R$. For instance, the interaction between two charges scales with $1/R$ and dipole-dipole interactions with $1/R^3$. More generally, when we label the two multipoles with ranks $l_A$ and $l_B$, then the interaction between two multipoles scales as $1/R^{l_A+l_B+1}$. From here on, we only consider neutral molecules such that dipole-dipole is the dominant contribution.

The electrostatic interaction depends on the relative orientation of the molecules \cite{avoird:80}.
For example, the dipole-dipole interaction is attractive for head-to-tail orientations, whereas it is repulsive for side-to-side approach.
Orienting or aligning molecules in space requires external fields \cite{friedrich:91,friedrich:91a}.
Without external fields, molecules are allowed to rotate freely such that in the frame of the lab their average dipole moment is zero.
This means that for molecules in pure rotational states, the first-order dipole-dipole interaction vanishes.
One exception to this is when two molecules are in rotational states that differ by one quantum.
In this case the rotational excitation can reside on either molecule, and the dipole-dipole interaction couples these two states leading to
a \emph{resonant} dipole-dipole interaction \cite{mason:62}.

The dipole-dipole interaction can couple to excited states of the molecule-molecule dimer,
which in second-order perturbation theory gives rise to a van der Waals interaction given by $-C_6/R^6$ \cite{margenau:39} with van der Waals coefficient $C_6$.
We can distinguish between the contributions of virtual excitations to electronically excited states, $C_6^\mathrm{elec}$,
and contributions of virtual excitations to different rotational states of both molecules within the electronic ground state, $C_6^\mathrm{rot}$ \cite{kotochigova:10}.
The electronic contribution is what is typically thought of as the van der Waals interaction \cite{stone:96}.
For many bialkali molecules the rotational contribution is two orders of magnitude larger \cite{zuchowski:13},
due to the small rotational excitation energy (0.1~cm$^{-1}$) compared to electronic ones (10\,000~cm$^{-1}$).
The electronic contribution to the van der Waals interaction is almost always attractive, which is a result of the spectrum of electronic states which generally becomes denser for more highly excited states.
We show that while the rotational van der Waals attraction is also attractive for two molecules in the same rotational state,
the interaction between molecules in different rotational states is always repulsive.
This results directly from the structure of the rotational energy levels which --unlike the electronic spectrum-- becomes sparser for higher excitations.

In a thermal gas of molecules, many different rotational states may be populated and collisions between molecules happen at energies higher than rotational energies.
In such a situation, the collision dynamics is determined by interactions at relatively short distances,
where they become comparable to the collision energy, and hence are strong compared to the energy difference between rotational states.
Therefore one typically does not think of the interaction as first and second-order perturbative couplings between pairs of rotational states \cite{stone:96}.
However, it has become possible to prepare molecules in a single rotational state and at ultracold temperatures of around a microkelvin while having precise control over the molecules using external fields \cite{fitch:21,softley:23,kaufman:21,liu:22}.
At these incredibly low temperatures and collision energies, the collision dynamics is determined by long-range interactions.
Of particular interest are molecules that posses a sizeable permanent dipole moment (e.g., NaK, KCs).
For these molecules the rotational contribution to the van der Waals interaction is by far dominant over the electronic one \cite{zuchowski:13}.

In order to create stable quantum gases of molecules that can be used as a platform for quantum technologies, one needs to suppress the loss of molecules due to collisions \cite{takekoshi:14,voges:20,guo:18,ye:18,gregory:19,cheuk:20}.
Even for chemically nonreactive molecules fast losses are observed, which could result from long-lived complex formation \cite{mayle:12,mayle:13,christianen:19a}.
Shielding molecules from collisions by creating repulsive interaction potentials has been realized using static electric fields \cite{quemener:16,gonzalez:17,matsuda:20,li:21} and microwaves \cite{karman:18d,karman:19c,karman:20,anderegg:21,schindewolf:22}.
We show in this paper that loss can also be suppressed in collisions between two molecules in different rotational states without the need of external field control.
We consider exploiting these interactions in several applications.
First, an exciting idea is to leverage these interaction towards efficient forced evaporation of a fermi mixture of molecules in different rotational states.
This intuitively appears possible since collisions between molecules in the same rotational state are suppressed by the $p$-wave centrifugal barrier,
and loss in $s$-wave collisions between distinguishable molecules in different rotational states is strongly suppressed by the repulsive interactions,
while these still lead to elastic collisions and thermalization.
Unfortunately we find that realistically the ratio of good-to-bad collisions remains limited.
Promising applications that we consider are in suppressing loss in on-site encounters of fermionic molecules in an optical lattice for quantum simulation,
as well as studying impurity physics with controlled dipolar interactions while suppressing collisional loss.

The structure of this paper is as follows: Section~\ref{sec:molmol} discusses qualitatively the dipole-dipole interaction between molecules in different rotational states in more detail, quantitatively calculates strength of the rotational van der Waals interaction, describes higher multipoles as well as the electronic contribution to the van der Waals interaction, and characterizes pairs of rotational states that are degenerate for which the collision dynamics is vastly different; Section~\ref{sec:collisions} provides computational details, illustrates interaction potentials for NaK and KCs for various rotational pairs, computes collisional loss rates by means of quantum scattering calculations, shows that a dipole can still be induced using static electric fields and microwaves, and explains the loss rates observed for degenerate rotational pairs; Section \ref{sec:applications} describes potential applications utilizing repulsive interactions for the evaporation of fermi mixtures, fermions in optical lattices and impurity physics; and Section~\ref{sec:conclusion} concludes this paper.

\section{Molecule-molecule interactions \label{sec:molmol}}

\subsection{Dipole-dipole interaction \label{sec:dipdip}}

The leading term in the multipole expansion of the electrostatic interaction between two neutral molecules is the dipole-dipole interaction. Dipolar interactions only couple rotational states $j$ and $j\pm1$. For a pair of molecules, this selection rule holds for each molecule separately. In second-order perturbation theory, the interaction strength is determined by virtual dipole-dipole \mbox{(de-)excitation} to other states. This strength is proportional to the square of the coupling and inversely proportional to the energy difference between the states, see Eq. \eqref{eq:secondorder} for more detail. This means that the interaction is dominated by coupling to states closest in energy and that the sign of this energy difference determines whether the interaction is attractive or repulsive. 

The rotational energy ladder is given by $E_j=Bj(j+1)$, with rotational constant $B$. In contrast to electronic states, where the excited levels become more closely spaced, excited rotational levels become sparser. This unique structure results in attractive interactions when $\Delta j=0$ and repulsive ones when $\Delta j\geq 2$, where $\Delta j=|j_A-j_B|$ denotes the difference in rotational states between the two molecules. We discuss the separate cases in more detail.

The van der Waals interaction is always attractive for two molecules in the same rotational state $j$. This is easiest to see when both molecules are in their rotational ground state. When $j_A=j_B=0$, the only possible second-order coupling is to a state higher in energy, where both molecules are virtually excited to $j=1$. This results in an attractive interaction where $C_6^\mathrm{rot}=d^4/6B$ with dipole moment $d$.
If both are in the same $j>0$ state, there are also contributions from de-excitation of one or both molecules. Excitation of a molecule always yields a larger energy difference $E_{j+1}-E_j=2B(j+1)$ than de-excitation $E_j-E_{j-1}=2Bj$ due to the structure of the rotational energy ladder. The pair state that lies closest to the $j_A=j_B=j$ state is the one where $j_A$ is excited and $j_B$ is de-excited, or vice versa, which is higher in energy by $2B$, e.g., $E_1+E_1<E_0+E_2$. In second order, this leads to an attractive van der Waals interaction when both molecules are in the same rotational state.

When $\Delta j=1$, e.g., the pair state $j_A+j_B=0+1$, they experience dipole-dipole interactions in first order due to resonant interactions with $1+0$. This interaction scales with $R^{-3}$ and results in fast collisions. 

\begin{figure}
    \centering
    \includegraphics[width=0.95\linewidth]{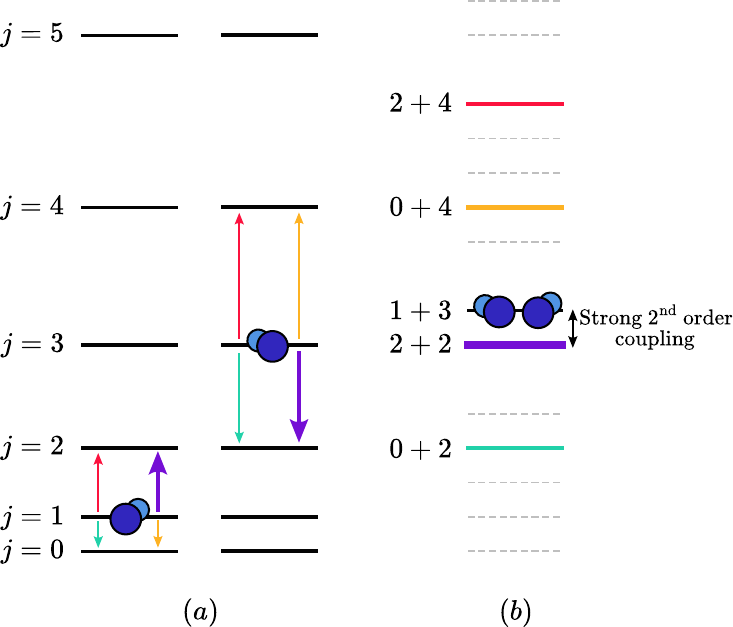}    
    \caption{\textbf{Schematic representation of rotational energy levels} for (a) two molecules in different rotational states (here, $j=1$ and $j=3$) and (b) the combined pair (here, $j+j'=1+3$). Different colors represent the various ways of virtually (de-)exciting the molecules via dipole-dipole interactions, creating strong second-order coupling to a state barely below the original state.}
    \label{fig:rep_diagram}
\end{figure}

For $\Delta j\geq 2$, the interaction is again second order.
Since the molecules are in different rotational states, it now matters whether one excites the one in the lower rotational state $j_<$ and de-excite the higher rotational state $j_>$, or vice versa. The smallest energy difference is found to be excitation of the $j_<$ molecule and de-excitation of $j_>$. This state lies \textit{below} the state of interest by $\Delta E=-2B(j_>-j_<-1)$. Figure~\ref{fig:rep_diagram} shows this schematically. In second order, this therefore results in a \textit{repulsive} interaction.

In addition to this qualitative understanding, we quantitatively determine the rotational van der Waals coefficients, $C_6^\mathrm{rot}$, for various pairs of rotational states.
The dipole-dipole interaction between two molecules is given by
\begin{equation}
    \hat{V}_\mathrm{dd}(R)=\frac{1}{4\pi\epsilon_0R^3}\left(\hat{\bm{d}}^{(A)}\cdot\hat{\bm{d}}^{(B)}-3\hat{d}_z^{(A)}\hat{d}_z^{(B)}\right)\,,
    \label{eq:Vdd}
\end{equation}
in the body-fixed frame where the intermolecular axis lies along the $z$ axis.
The $C_6^\mathrm{rot}$ coefficient between rotational states $j_A$ and $j_B$ can then be calculated using second-order perturbation theory as
\begin{equation}
    \sum_{n'\neq n}\frac{\left|\langle n'|\hat{V}_\mathrm{dd}(R)|n\rangle\right|^2}{E_n-E_{n'}}=-\frac{C_6^\mathrm{rot}}{R^6}.
    \label{eq:secondorder}
\end{equation}
In practice we used a space-fixed basis set with functions $|n\rangle=|(j_A,j_B),j_{AB},L; J M p\rangle$ and energies $E_n=B[j_A(j_A+1)+j_B(j_B+1)]$,
see Sec.~\ref{sec:computational} for a detailed description.
Here, the quantum number $j_{AB}$ represents the coupled angular momentum of $j_A$ and $j_B$, $L$ the angular momentum number associated with end-over-end rotation, $J$ the quantum number of the total angular momentum, $M$ the total angular momentum along a space-fixed quantization axis, and $p=\pm1$ corresponds with the permutation symmetry of bosons ($+1$) and fermions ($-1$) such that if $j_A=j_B$ only basis functions where $p=(-1)^{j_A+j_B-j_{AB}+L}$ contribute.
Since $j_A=j_B$ channels couple by dipole-dipole only to channels with $|j_A-j_B|=0$ or $2$,
the differences between bosons and fermions disappear for $\Delta j > 2$.
The electrostatic interactions are diagonal in $J$, $M$, $p$, and are independent of $M$.
For $L=0$ channels, $j_{AB}=J$ and hence $j_{AB}$ is effectively a good quantum number within the $s$-wave channels.

\begin{table}
    \centering
    \caption{\textbf{Rotational van der Waals $C_6^\mathrm{rot}$ coefficients} for pairs of bosonic and fermionic molecules in various rotational states $j_A$ and $j_B$, where $j_A\geq j_B$, in units of $d^4/B$. Different values per pair correspond to various $j_{AB}$ in increasing order.}
    \begin{tabular}{cccccccc}
        \hline\hline
        &  & \multicolumn{6}{c}{$j_A$} \\ 
        &  & \multicolumn{2}{c}{0} & \multicolumn{2}{c}{1} & \multicolumn{2}{c}{2} \\ 
        \cmidrule(lr){3-4}\cmidrule(lr){5-6}\cmidrule(lr){7-8}
        &  & bosons & fermions & bosons & fermions & bosons & fermions \\ \hline
        \multirow{15}{*}{$j_B$} & 0 & $+0.167$ & - &  &  &  & \\
        & 1 & - & - & \begin{tabular}[c]{@{}c@{}}$+0.290$\\$+0.104$\end{tabular} & $+0.230$ &  & \\
        &   &          &          &   &   & $+0.298$ & \multirow{3}{*}{\begin{tabular}[c]{@{}c@{}}$+0.274$\\$+0.163$\end{tabular}} \\
        & 2 & $-0.097$ & $-0.070$ & - & - & $+0.228$ & \\
        &   &          &          &   &   & $+0.085$ & \\
        &   &          &          & $-0.095$ & $-0.093$ &  & \\
        & 3 & $-0.033$ & $-0.033$ & $-0.062$ & $-0.076$ & - & - \\
        &   &          &          & $-0.062$ & $-0.027$ &  & \\
        &   &          &          &          &          & $-0.097$ & $-0.097$ \\
        &   &          &          & $-0.036$ & $-0.036$ & $-0.083$ & $-0.086$ \\
        & 4 & $-0.019$ & $-0.019$ & $-0.027$ & $-0.027$ & $-0.074$ & $-0.065$ \\
        &   &          &          & $-0.019$ & $-0.019$ & $-0.042$ & $-0.061$ \\
        &   &          &          &          &          & $-0.053$ & $-0.014$ \\ 
        \hline\hline
    \end{tabular}
    \label{tab:C6_jajb}
\end{table}

\begin{figure}
    \centering
    \includegraphics[width=\linewidth]{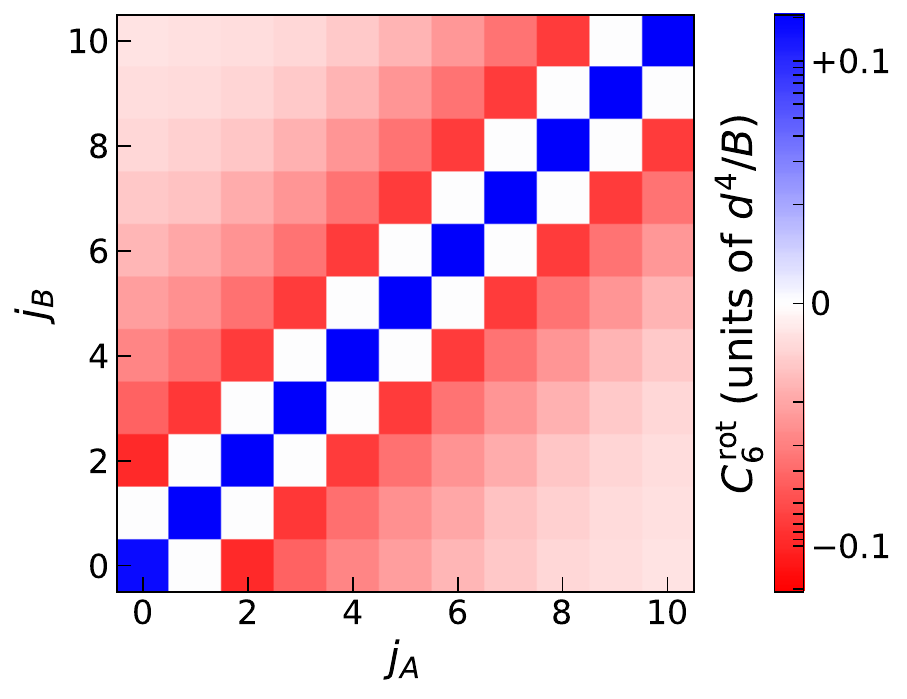}
    \caption{\textbf{Rotational van der Waals $C_6^\mathrm{rot}$ coefficients} on a logarithmic scale for pairs of bosonic molecules in various rotational states $j_A$ and $j_B$. For each pair, an average is taken over the various ways of coupling both together.}
    \label{fig:C6_jajb}
\end{figure}

We calculate $C_6^\mathrm{rot}$ for the $L=0$ states, although in the basis we also include $L=2$ to account for dipole-dipole coupling. The values for different pairs of $j_A$ and $j_B$ in units of $d^4/B$ are given in Table \ref{tab:C6_jajb} and a visual representation is shown in Fig.~\ref{fig:C6_jajb}.
Due to the anisotropy of the dipole-dipole interaction,
the second-order interaction depends on $j_{AB}$ for $j_A,j_B>0$.
In the figure we have averaged over the various possible ways of coupling to $j_{AB}$ for each pair and only show results for bosonic molecules, although the case of fermionic molecules is qualitatively very similar. Moreover, since the dominant interaction for $\Delta j=1$ is first order, not second order, these values are not given.

From Table \ref{tab:C6_jajb} and Fig.~\ref{fig:C6_jajb} we can clearly see the attractive van der Waals interaction when $j_A=j_B$.
For all other pairs where $\Delta j\geq 2$ the interaction is repulsive. This decreases in strength as $\Delta j$ increases, since the energy difference between the prepared and virtually (de-)excited states increases.

\subsection{Higher multipoles and electronic contribution \label{sec:multipoles}}

Higher-order multipole-multipole interactions also play a role. For a multipole of rank $l$, the selection rules that couple a rotational state $j$ to some state $k$ are given by the the triangular condition $|j-k|\leq l\leq j+k$ and the parity rule that $j+k+l$ must be even. Therefore, if both \mbox{$j,k>0$}, there are various multipoles that contribute to the full electrostatic interaction. For multipoles of rank $l_A$ and $l_B$ this contribution scales as $R^{-l_A-l_B-1}$. The dipole-dipole interaction that scales as $R^{-3}$ is therefore typically dominant. In special cases of degeneracy, see Secs. \ref{sec:degeneracy} and \ref{sec:degenerate_calcs}, we find that, for example, dipole-quadrupole can occur in first order. In other cases, including higher multipoles affects the results only quantitatively.

At first glance, it would seem that for states where both $j_A,j_B>0$ quadrupole-quadrupole coupling is possible in first order, which would scale as $R^{-5}$. However, in the $s$-wave channel this anisotropic interaction has zero expectation value and the interaction couples the $s$-wave to the $g$-wave channel such that in second order the interaction only scales as $R^{-8}$.

Besides the electrostatic contribution to the van der Waals interaction, there is still the electronic contribution. For bialkalis with a sizeable dipole moment, typical electronic and rotational contributions to the van der Waals coefficients are $C_6^\mathrm{elec}\sim10^4$ a.u. and $C_6^\mathrm{rot}\sim d^4/B\sim10^6$ a.u., such that the electronic contribution is generally two orders of magnitude smaller, although still present \cite{zuchowski:13}. This holds for molecules such as NaK and KCs, but not for KRb due to its small dipole moment. In this work we include this as an isotropic interaction that is the same for all rotational states. When $\Delta j\geq 2$, the strength of the repulsive interaction decreases for larger $\Delta j$, while the electronic contribution remains constant. Therefore, for larger $\Delta j$, the electronic contribution is important and can completely suppress the repulsive part of the rotational contribution.

\subsection{Pair degeneracies \label{sec:degeneracy}}

For a pair of linear molecules of the same species, the combined states $j+j'$ can become degenerate with another pair of $k+k'$ rotational states. When this is the case, the coupling between the two states does not occur in second order, but in first order, which drastically changes the interaction and resulting collision rates. Here, we derive which pairs of rotational pair states are degenerate, and we discuss the physically most relevant degeneracies.

If two molecules are the same species, i.e., both have the same rotational constant $B$, then degeneracy, which we denote by $j+j'\equiv k+k'$, holds if $E_j+E_{j'}=E_k+E_{k'}$, which gives
\begin{equation}
    j(j+1)+j'(j'+1)=k(k+1)+k'(k'+1)
    \label{eq:deg}
\end{equation}
for which we can find integer solutions. By completing squares and using $a^2-b^2=(a-b)(a+b)$ one arrives at
\begin{equation}
    (j-k)(j+k+1)=(k'-j')(k'+j'+1)\,.
    \label{eq:deg_eq}
\end{equation}
We then define $l=j-k$ and $l'=k'-j'$, which correspond to the lowest rank multipoles that couple the rotational states. We assume without loss of generality that $j>k$ and $k'>j'$.
Both sides of Eq.~\eqref{eq:deg_eq} equal the same integer, where the left-hand side is divisible by $l$ and the right-hand side by $l'$. This means that any factors that divide $l$ but not $l'$, should divide $k'+j'+1$. Equation~\eqref{eq:deg_eq} must therefore equal an integer of the form $nll'/\mathrm{gcd}(l,l')$, where $n$ is any positive integer and $\mathrm{gcd}(l,l')$ the greatest common divisor of $l$ and $l'$. This leads to $j+k+1=nl'/\mathrm{gcd}(l,l')$ and $k'+j'+1=nl/\mathrm{gcd}(l,l')$, and together with the definitions of $l$ and $l'$ this yields
\begin{equation}
\begin{aligned}
    j&=k+l\,,\\
    j'&=\frac{1}{2}\left(\frac{nl}{\mathrm{gcd}(l,l')}-l'-1\right)\,,\\
    k&=\frac{1}{2}\left(\frac{nl'}{\mathrm{gcd}(l,l')}-l-1\right)\,,\\
    k'&=j'+l'\,.
\end{aligned}
\end{equation}
These only give integer solutions if
\begin{enumerate}
    \item both $l,l'$ are odd and $n$ is even, or
    \item $l$ is odd, $l'$ is even (or vice versa) and $n$ is odd, or
    \item both $l,l'$ are even, containing the same power of 2 in their prime factorization, and $n$ is odd.
\end{enumerate}
The last requirement shows, for example, that for $l=2$ and $l'=4$ there are no degeneracies, while for $l=2$ and $l'=6$ there are. Positivity of all quantum numbers is ensured in both cases if $n\geq\mathrm{gcd}(l,l')\max(\frac{l+1}{l'},\frac{l'+1}{l})$ and $l,l'\geq1$. Given $l$ and $l'$ as restricted above, one can then generate a family of degenerate states parametrized by $n$.

Equation~\eqref{eq:deg} can be extended to higher degrees of degeneracies, where finding different degenerate pairs becomes mathematically equivalent to finding the different ways of writing an integer as a sum of two squares of odd numbers. For example, for a three-fold degeneracy, the lowest integer for which this holds is 650, resulting in $8+9\equiv 5+11\equiv 2+12$.

We find all degenerate pairs that are coupled by dipole-dipole interactions, where $l=l'=1$. 
From the recipe above these are given by
\begin{align*}
    1+0&\equiv0+1\,,\\
    2+1&\equiv1+2\,,\\
    3+2&\equiv2+3\,,\text{  etc.}
\end{align*}
In general, for $l=l'$, one finds the trivial solutions $j+j'\equiv j'+j$,
which correspond to the known resonant multipole-multipole interactions. 

For dipole-quadrupole coupling, where $l=1$ and $l'=2$, we find non-trivial degenerate pairs. Following the recipe above, we must take $n=3,5,7,...$ . The degenerate pairs are then of the form $n+\frac{n-3}{2}\equiv\frac{n+1}{2}+(n-1)$. So, for increasing $n$ we find the degenerate pairs 
\begin{align*}
    3+0&\equiv2+2\,,\\
    5+1&\equiv3+4\,,\\
    7+2&\equiv4+6\,,\text{  etc.}
\end{align*}

Exact degeneracies as described above only hold for rigid rotors, where no centrifugal distortion is taken into account. In reality, centrifugal forces pull the atoms apart and the usual equation for the rigid rotor energy levels is extended to
\begin{equation}
    E_j=Bj(j+1)-Dj^2(j+1)^2\,,
\end{equation}
with $D\ll B$ the centrifugal distortion constant. This splits the `degenerate' states by a small amount compared to the rotational level spacing. The interaction potentials then contain a crossing between the two coupled states, which can significantly change the dynamics of the collision compared to non-degenerate states.

Of special interest is the $3+0\equiv2+2$ degeneracy. The centrifugal distortion lifts this degeneracy such that the $3+0$ state is slightly lower in energy.
The dipole-dipole interaction leads to an attractive van der Waals interaction for the $2+2$ state and a repulsive van der Waals interaction for the $3+0$ state.
The dipole-quadrupole coupling between these \textit{nearly} degenerate states creates a repulsion between them, possibly stabilizing the upper $2+2$ state and destabilizing the $3+0$ state. We explore this idea quantitatively in section \ref{sec:degenerate_calcs}.
 
\section{Interaction potentials and collisions \label{sec:collisions}}

\subsection{Computational details \label{sec:computational}}

The colliding molecules $X=A,B$ are modeled as rigid rotors with monomer Hamiltonian
\begin{equation}
    \hat{H}^{(X)}=B\left[\hat{j}^{(X)}\right]^2-D \left[ \hat{j}^{(X)}\right]^4 -\hat{\bm{d}}^{(X)}\cdot\bm{E}_\text{static}+\hat{H}_\text{ac}^{(X)}\,,
    \label{eq:monomerH}
\end{equation}
where the first term describes the rotational kinetic energy, the second term the centrifugal distortion, the third the Stark interaction between the molecular dipoles and a static electric field, and the fourth the interaction with a microwave field. These last two terms are only included in those calculations where a static electric field or a microwave field is used. For a more detailed description of this Hamiltonian and subsequent scattering calculations we refer to the Supplemental Material of Ref. \cite{karman:18d}. Note that we have not included hyperfine structure in these calculations as the hyperfine energy scale is negligible compared to rotational energy-level structure.

The total dimer Hamiltonian is given by
\begin{align}
    \hat{H}=&-\frac{\hbar^2}{2\mu}\frac{1}{R}\frac{\mathrm{d}^2}{\mathrm{d}R^2}R+\frac{\hat{L}^2}{2\mu R^2}+\hat{H}^{(A)}+\hat{H}^{(B)}\nonumber\\
    &+\hat{V}_\mathrm{dd}(R)+\hat{V}_\mathrm{dq}(R)+\hat{V}_\mathrm{qq}(R)-\frac{C_6^\mathrm{elec}}{R^6}\,,
    \label{eq:dimerH}
\end{align}
where $\mu$ is the reduced mass, $R$ the intermolecular distance, $\hat{L}$ the angular momentum operator associated with the end-over-end rotation, and the dipole-dipole, dipole-quadrupole, and quadrupole-quadrupole interactions given by
\begin{align}
    \hat{V}_\mathrm{dd} &= -\frac{\sqrt{30} d^2}{R^3} \left[ \left[ C_{1}\left(\hat{r}^{(A)} \right) \otimes C_{1}\left(\hat{r}^{(B)} \right)  \right]^{(2)} \otimes C_{2}\left(\hat{R}\right) \right]^{(0)}_0, \nonumber \\
    \hat{V}_\mathrm{dq} &= \frac{\sqrt{105} dQ}{R^4} \Bigg\{ \left[ \left[ C_{1}\left(\hat{r}^{(A)} \right) \otimes C_{2}\left(\hat{r}^{(B)} \right)  \right]^{(3)} \otimes C_{3}\left(\hat{R}\right) \right]^{(0)}_0 \nonumber \\ -& \left[ \left[ C_{2}\left(\hat{r}^{(A)} \right) \otimes C_{1}\left(\hat{r}^{(B)} \right)  \right]^{(3)} \otimes C_{3}\left(\hat{R}\right) \right]^{(0)}_0 \Bigg\}, \nonumber \\
    \hat{V}_\mathrm{qq} &= \frac{\sqrt{630} Q^2}{R^5} \left[ \left[ C_{2}\left(\hat{r}^{(A)} \right) \otimes C_{2}\left(\hat{r}^{(B)} \right)  \right]^{(4)} \otimes C_{4}\left(\hat{R}\right) \right]^{(0)}_0,
\end{align}
where $Q$ is the quadrupole moment, $C_l(\hat{r})$ is the rank $l$ tensor whose spherical components are the Racah normalized spherical harmonics depending on the polar angles of the vector $\hat{r}$,
and 
\begin{align}
    \left[\hat{A}_{k_A} \otimes \hat{B}_{k_B}\right]^{(k)}_q = \sum_{q_A,q_B} \hat{A}_{k_A, q_A} \hat{B}_{k_B,q_B} \langle k_A q_A k_B q_B | k q\rangle
\end{align}
is the $q$ spherical component of the rank $k$ irreducible tensor product of the rank $k_A$ spherical tensor $\hat{A}$ and the rank $k_B$ spherical tensor $\hat{B}$.

We make use of an uncoupled basis set, where we express the basis functions of the monomers using angular momentum states $|jm_j\rangle$. The end-over-end rotation of the molecules about one another is described by the state $|LM_L\rangle$. Lastly, in the case where we include microwaves, the basis also contains a photon state $|N\rangle$ representing the number of photons taken up by the complex. The full dimer basis functions are then of the form
\begin{equation}
    |j^{(A)}m^{(A)}\rangle|j^{(B)}m^{(B)}\rangle|LM_L\rangle |N\rangle.
    \label{eq:basis}
\end{equation}
The resulting basis can be reduced by adapting it to permutation symmetry and restricting good quantum numbers \cite{karman:18d}. 
We perform separate calculations for the two parities of $(-1)^{j_A+j_B+L}$ as well as for different values of $M_\mathrm{tot}=m^{(A)}+m^{(B)}+M_L$. For convergence, we have used a basis up to $L_\mathrm{max}=7$, $|M_{L,\mathrm{in}}|_\mathrm{max}=1$ and $j_\mathrm{max}=\max(j^{(A)}_\mathrm{in},j^{(B)}_\mathrm{in})+2$ for all calculations, with `in' representing the initially prepared rotational state. For states where $\Delta j=1$, we have converged the results up to $|M_{L,\mathrm{in}}|_\mathrm{max}=5$ due to the strong resonant coupling. When microwaves are included the basis also contains $N=0$ and $-1$. For the results including electric fields or microwaves results are converged by including up to $j_\mathrm{max}=3$ and $|M_{L,\mathrm{in}}|_\mathrm{max}=3$. In all cases, results are converged within 1 \%.

Using the Hamiltonian as defined in Eqs. \eqref{eq:monomerH} and \eqref{eq:dimerH} and the dimer basis set \eqref{eq:basis}, we calculate loss rates by coupled-channels scattering calculations using the renormalized Numerov method \cite{johnson:78}. We propagate two sets of linearly independent solutions on a radial grid. At short range we impose absorbing boundary conditions to model the short-range losses of ultracold molecules and at long range $S$-matrix boundary conditions \cite{janssen:13ref}. The absorbing boundary assumes that at the boundary there is only flux going into short-range reactive channels, leading to 100 \% loss for these channels, as observed experimentally \cite{takekoshi:14,voges:20,guo:18,ye:18,gregory:19,cheuk:20}. At short enough distances our results are insensitive to the exact placement of the boundary condition. Calculations were converged with respect to the radial grid, which ranges from 50 to 10\,000 $a_0$ with at least 10 points per local de Broglie wavelength. After propagation, this results in a scattering $S$-matrix from which we get all state-to-state scattering probabilities. Collisional loss depends on the probabilities of short range and inelastic collisions. The cross sections for these loss mechanisms follow from the $S$-matrix as
\begin{align}
    \sigma_\mathrm{short}&=\frac{\pi}{k^2}\sum_{r,L,M_L}\left|T_{r;i,L,M_L}^\mathrm{(short)}\right|^2\,,\\
    \sigma_\mathrm{inel}&=\frac{\pi}{k^2}\sum_{f\neq i,L',M_L',L,M_L}\left|T_{f,L',M_L';i,L,M_L}^\mathrm{(long)}\right|^2\,,
\end{align}
where $\bm{T}=\bm{1}-\bm{S}$ and $k=\sqrt{2\mu E}$. The label $r$ stands for all adiabatic channels that are classically allowed at the capture distance and the labels $i$ and $f$ for the initial and final rotational states of the molecules. When both molecules are in the same $|jm_j\rangle$ state, they are indistinguishable and the cross sections should be multiplied by 2. The energy-dependent cross-section, $\sigma(E)=\sigma_\mathrm{short}+\sigma_\mathrm{inel}$, can then be averaged over a Maxwell-Boltzmann distribution for a given temperature $T$ to yield thermal rate coefficients
\begin{equation}
    k_2=\sqrt{\frac{8k_BT}{\pi\mu}}\frac{1}{(k_BT)^2}\int_0^\infty\sigma(E)\exp\left(-\frac{E}{k_BT}\right)E\;\mathrm{d}E\,.
\end{equation}
We integrate using 7 logarithmically spaced collision energies between $0.1T$ and $10T$. All results in this paper are given for $T=1$ $\mu$K.

The molecules that we study here are NaK and KCs \cite{voges:20,park:15,grobner:17}. Both have a large dipole moment \cite{gerdes:11,igelmann:86}, a small electronic contribution to the long-range interaction \cite{zuchowski:13}, and both are non-reactive \cite{zuchowski:10b}. Moreover, due to the available isotopes of K, both bosonic and fermionic NaK and KCs can be produced. Note that the theory presented above applies generally for other dipolar, linear molecules for which the rotational contribution to the van der Waals interaction is dominant. The molecular parameters used for NaK are $B=9.41\times10^{-2}$ cm$^{-1}$, $D=8\times10^{-7}$ cm$^{-1}$ \cite{allegrini:77}, $d=2.72$ Debye \cite{gerdes:11}, $Q=10.6$ a.u. \cite{christianen:19} and $C_6^\mathrm{elec}=8.28\times10^{3}$ a.u. \cite{zuchowski:13}, and for KCs we used $B=3.00\times10^{-2}$ cm$^{-1}$, $D=2.4\times10^{-8}$ cm$^{-1}$ \cite{ferber:09}, $d=1.92$ Debye \cite{igelmann:86}, $Q=6.1$ a.u. and $C_6^\mathrm{elec}=1.84\times10^{4}$ a.u. \cite{zuchowski:13}, where for KCs we have approximated $Q\approx dR_e$ with $R_e$ the equilibrium bond distance.

\subsection{Interaction potentials \label{sec:potentials}}

\begin{figure*}
    \centering
    \includegraphics[width=\linewidth]{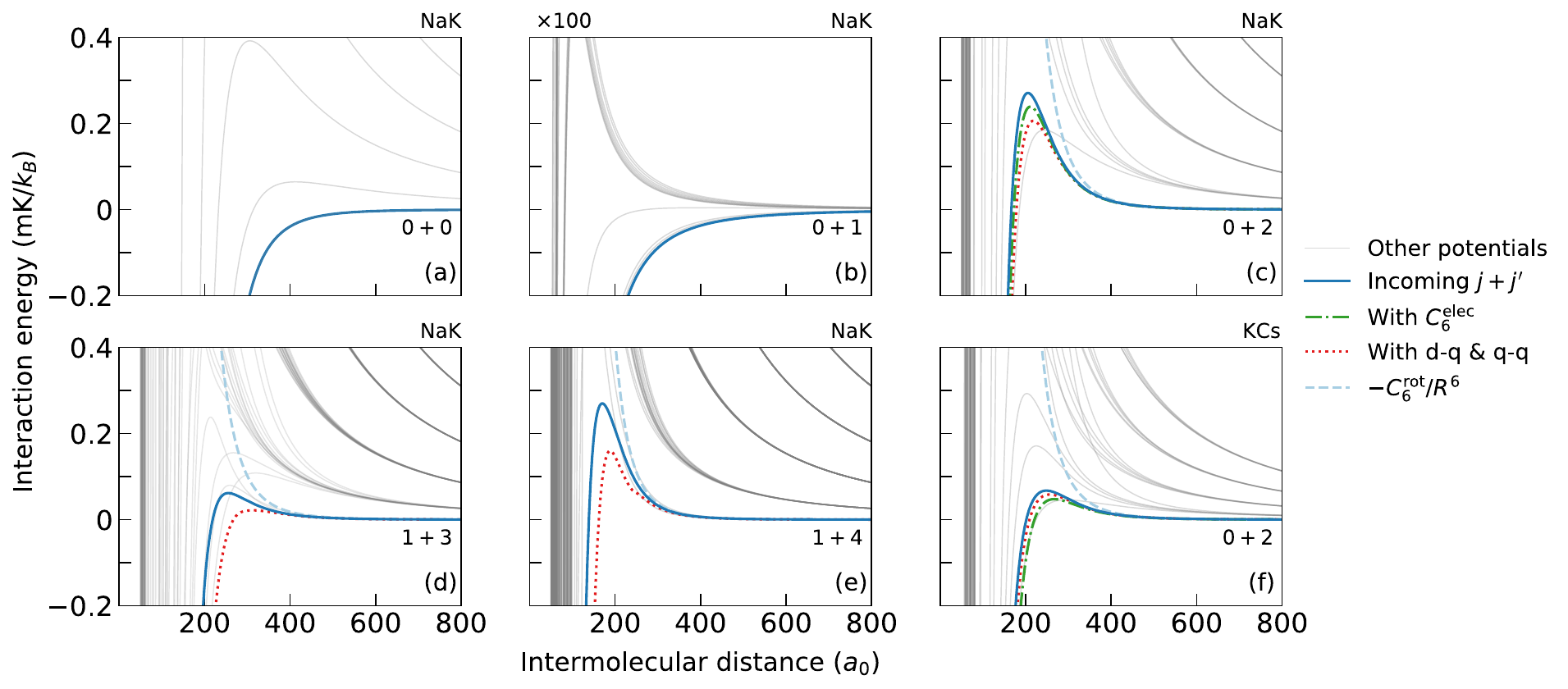}
    \caption{\textbf{Interaction potential curves} for bosonic (a) NaK $0+0$, (b) NaK $0+1$, (c) NaK $0+2$, (d) NaK $1+3$, (e) NaK $1+4$ and (f) KCs $0+2$, with dimer rotational states labeled as $j+j'$ where $m_j=m_{j'}=0$. All potentials are referenced to the energy of the initial state (solid blue lines) at infinite separation. Due to strong first-order interactions for $0+1$, the energy is scaled by 100 in panel (b). The pure second-order rotational interaction curve is given by the dashed light blue curve. Addition of an isotropic electronic $C_6^\mathrm{elec}$ interaction is given for (c) and (f) by the dash-dotted green curve. Addition of a quadrupole moment for dipole-quadrupole and quadrupole-quadrupole interaction is represent in (c)-(f) by the dotted red curve.}
    \label{fig:reppot}
\end{figure*}

We find adiabatic interaction potentials by diagonalizing the dimer Hamiltonian \eqref{eq:dimerH}, excluding the radial kinetic energy term, at fixed intermolecular distances $R$. For a number of rotational state pairs for bosonic NaK and KCs these potential energy curves are shown in Fig.~\ref{fig:reppot}. In all panels, the solid blue curve corresponds to the incoming $j+j'$ state with $L=0$, whereas the thin gray curves correspond to higher $L$, or different couplings between $j$ and $j'$ that have higher barriers. These solid lines are calculated including only the dipole-dipole interaction but excluding the dipole-quadrupole, quadrupole-quadrupole and electronic terms.

With both molecules in the ground rotational state ($j+j'=0+0$) in Fig.~\ref{fig:reppot}(a) we find the well-known attractive van der Waals interaction with $C_6^\mathrm{rot}=d^4/6B$. For the $0+1$ pair, panel (b) demonstrates the strong resonant $R^{-3}$ dipole-dipole interaction. Note that the energy scale for this panel has been extended by two orders of magnitude. In panels (c)-(e) we show the pairs $0+2$, $1+3$ and $1+4$, respectively, where $\Delta j\geq 2$. As discussed in section \ref{sec:molmol}, this results in a repulsive rotational van der Waals interaction at long range as shown by the dashed second-order $R^{-6}$ curve.
However, there is a distance below which the perturbative approach breaks down.
At very short distances, the molecules polarize one another, and ultimately the lowest interaction potentials become attractive.
At intermediate distances, this results in a potential barrier that can shield collisions at low enough collision energy.
The main loss mechanism at low collision energy then stems from tunneling through this barrier and reaching the short range, which depends on the height and width of the barrier as well as the mass of the colliding molecules. Panel (f) finally shows the $0+2$ pair for bosonic KCs, where a similar, though lower barrier is compared to NaK. 
Potential curves including the isotropic electronic van der Waals interaction are shown in panels (c)-(f) as the dash-dotted lines. This decreases the barrier heights and widths, though not by so much as to remove the barrier in these cases. For state pairs with larger $\Delta j$, where the repulsive $C_6^\mathrm{rot}$ becomes small compared to the attractive $C_6^\mathrm{elec}$, the barrier disappears and higher losses are again expected. 

In Figs.~\ref{fig:reppot}(c) and (f) we also include dipole-quadrupole and quadrupole-quadrupole interactions, which decrease the barrier width and height slightly.
This occurs even in cases where we expect no contribution to the $s$-wave potential from perturbation theory.
The electrostatic interaction cannot be described perturbatively at the position of the barrier. The various states then mix, where especially the rotational quantum numbers $j$ and $L$ cease to be good quantum numbers.Level repulsion between the various states lowers the lowest potential curve,
and lowers the height of the potential barrier of interest.

\subsection{Collisional loss rates \label{sec:loss}}

\begin{figure*}
    \centering
    \includegraphics[width=0.9\linewidth]{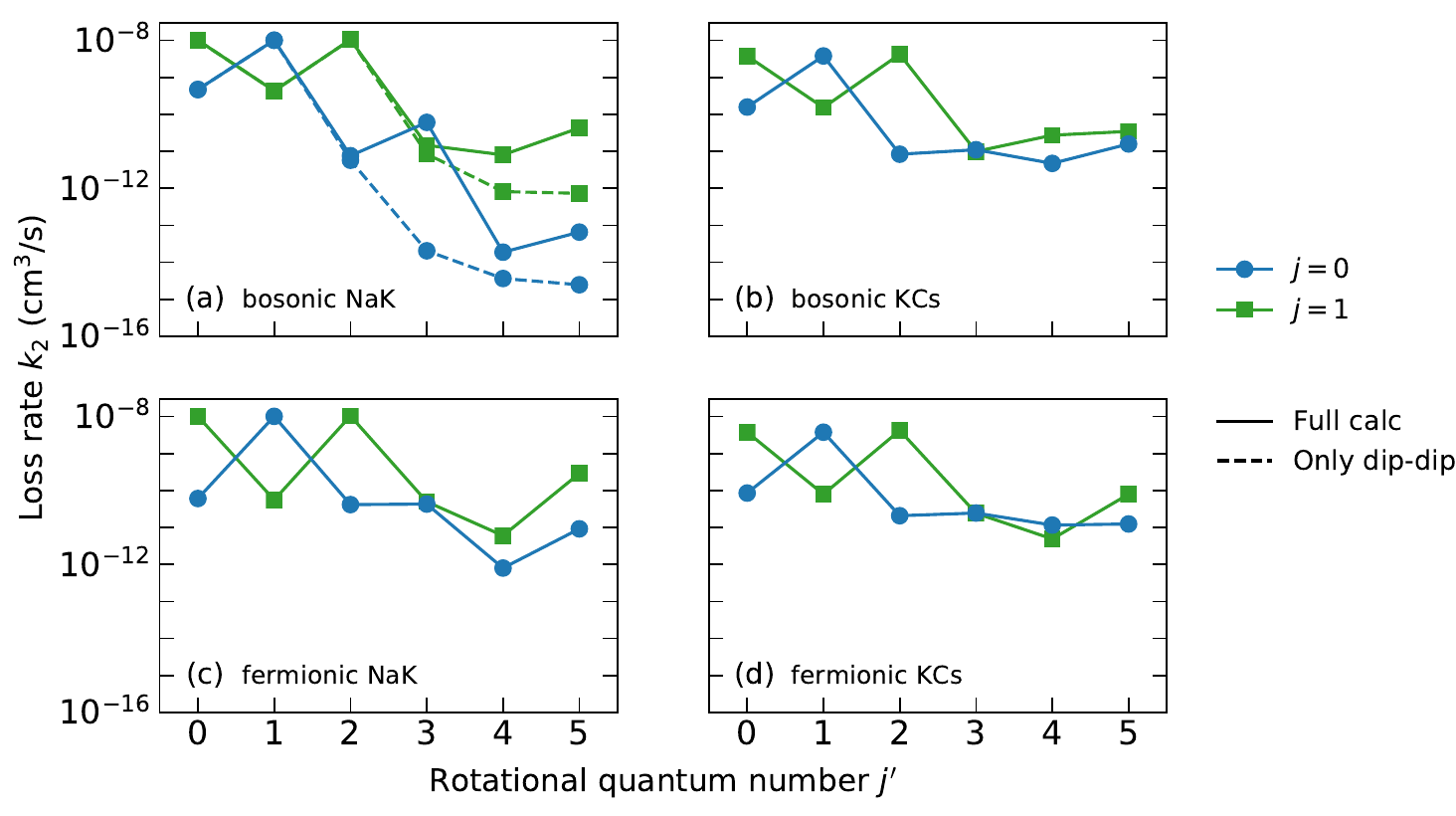}
    \caption{\textbf{Loss rates for various rotational state pairs} $j+j'$ with $m_j=m_{j'}=0$ for (a) bosonic NaK, (b) bosonic KCs, (c) fermionic NaK and (d) fermionic KCs. Panel (a) contains loss rates where only the dipole-dipole interaction is taken into account (dashed lines). The full calculations further include contributions from dipole-quadrupole, quadrupole-quadrupole, isotropic electronic $C_6^\mathrm{elec}$ and the centrifugal distortion constant.}
    \label{fig:loss}
\end{figure*}

We calculated loss rates for a range of rotational state pairs $j+j'$ with $m_j=m_{j'}=0$ for bosonic and fermionic species of NaK and KCs at 1 $\mu$K, which are shown in Fig.~\ref{fig:loss}. The dashed line in panel (a) only contains dipole-dipole interactions, whereas the full calculations in all panels further include dipole-quadrupole, quadrupole-quadrupole, isotropic electronic $C_6^\mathrm{elec}$ and centrifugal distortion terms. The last term is important for the degenerate $0+3(\equiv2+2)$ and $1+5(\equiv3+4)$ states due to strong dipole-quadrupole interactions. We explore this further in section \ref{sec:degenerate_calcs}.

For $0+0$ collisions we find loss rates close to universal loss \cite{idziaszek:10}, as expected. For bosonic and fermionic NaK, universal loss rates at 1 $\mu$K are $6\times10^{-10}$ and $6\times10^{-11}$ cm$^3$/s, respectively, and for KCs $3\times10^{-10}$ and $1\times10^{-10}$ cm$^3$/s. For fermions, the universal loss rate scales linearly with temperature due to the presence of the $p$-wave centrifugal barrier, and hence the rate is lower than for bosons. Similar loss rates are found for $1+1$ collisions. The highest loss rates we find when $\Delta j=1$ due to strong resonant dipole-dipole interactions, which holds here for $0+1\equiv1+0$ and $1+2\equiv 2+1$ pairs. For the states where $\Delta j\geq 2$ one can see from the dipole-dipole calculations that loss rates decrease rapidly by orders of magnitude as a function of $\Delta j$ as the attractive interactions occur at shorter distances, creating a higher and wider barrier.

Dipole-quadrupole, quadrupole-quadrupole and electronic terms lower the barriers or even cause them to disappear, especially at larger $\Delta j$, where the rotational contribution decreases. We find that the loss rates can be orders of magnitude smaller than universal loss. Since the barrier for KCs is generally lower than for NaK, loss rates for KCs at larger $\Delta j$ are higher than for NaK.

\begin{figure}
    \centering
    \includegraphics[width=\linewidth]{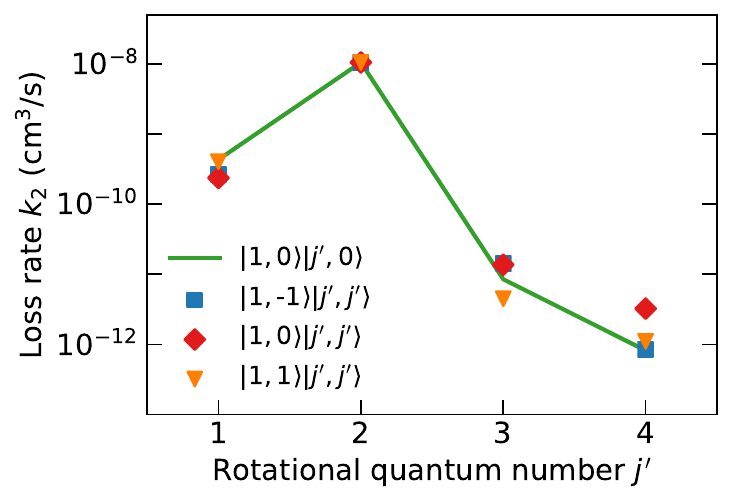}
    \caption{\textbf{Spread of loss rates} for rotational state pairs $1+j'$ for different $m_j$ and $m_{j'}$ for bosonic NaK. The green curve corresponds to the dashed green curve in Fig.~\ref{fig:loss}(a) and accounts only for the dipole-dipole interaction.}
    \label{fig:loss_m}
\end{figure}

The interaction between molecules is anisotropic for nonzero $j$ and $j'$, which leads to a dependence on $m_j$ and $m_j'$.
We explore this dependence by calculating loss rates for bosonic NaK $1+j'$ states with different $m_j$ and $m_{j'}$, as shown in Fig.~\ref{fig:loss_m}.
The solid line in Fig.~\ref{fig:loss_m} corresponds to the dashed line in Fig.~\ref{fig:loss}(a). The various markers then show the spread in loss rates, which are at most an order of magnitude, which still shows that irrespective of the choice of $m_j$ and $m_j'$, loss rates are suppressed for $\Delta j\geq 2$.

\begin{figure}
    \centering
    \includegraphics[width=\linewidth]{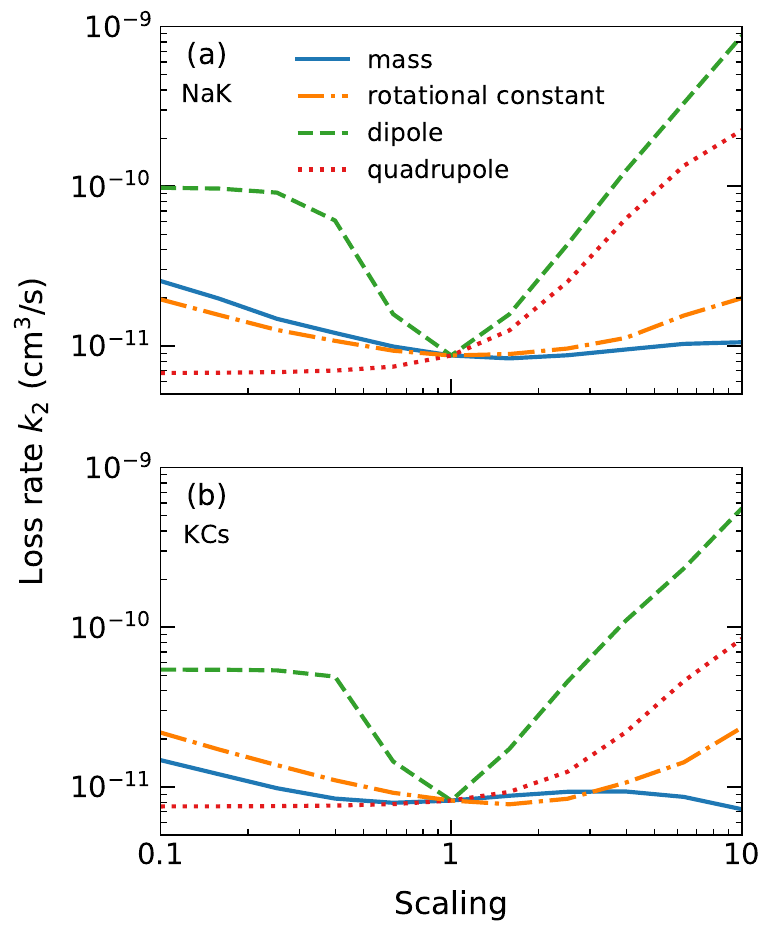}
    \caption{\textbf{Scaling of loss rates} for $j+j'=0+2$ of (a) bosonic NaK and (b) bosonic KCs with respect to mass, rotational constant, dipole moment and quadrupole moment. These results also include an isotropic $C_6^\mathrm{elec}$ contribution.}
    \label{fig:loss_scaling}
\end{figure}

We investigate the dependence on various molecular parameters by artificially scaling the mass, the rotational constant, the dipole moment and the quadrupole moment.
Results are given in Fig.~\ref{fig:loss_scaling},
where the two panels show results for NaK and KCs, respectively.
We note that both panels exhibit qualitatively similar behavior.
As the rotational constant is increased, the barrier shifts to shorter distances with decreasing width and increasing height, changing the loss rates only by a small factor. 
The effect of scaling of the dipole moment is most pronounced and shows that for both NaK and KCs the physical dipole moment results in smallest losses. Decreasing the dipole moment has the effect that the quadrupole couplings and the contribution from $C_6^\mathrm{elec}$ become relatively more important, such that losses increase. Below $\sim0.3d$ this has the effect that the loss rate plateaus as the barrier disappears.
Increasing the dipole moment increases the repulsive van der Waals interaction,
but also strengthens the attractive dipole-dipole interactions at short distances.
As a result, the barrier shifts to larger distances, resulting also in faster losses. 
Reducing the quadrupole moment has no effect as it does not determine the barrier.
When we increase the quadrupole moment, the quadrupole couplings between the potential curves become strong enough such that level repulsion lowers the barrier of the lowest adiabatic potential and increases loss rates.

Surprisingly, we find that the dependence on mass is relatively weak. There are actually two competing effects.
First, the loss is determined by the rate of tunneling through the repulsive potential,
and for a mass-independent potential one expects a steep suppression of the loss rate as the mass is increased. 
The other effect is that, perhaps surprisingly, even the $L=0$ potential is mass dependent at short enough distances.
As mentioned in the previous section, the barrier arises due to coupling by anisotropic interactions to other nearby states, including higher $L$ states. The centrifugal barrier for higher $L$ states depends on mass and when the perturbative treatment breaks down, the coupling to such states results in the barrier also becoming mass-dependent.
This is in stark contrast with the $s$-wave interaction potential for isotropic interactions, which is mass independent.
As mass increases, the barrier height decreases, resulting in increasing loss rates. The two effects are opposite and their near-cancellation results in the weak mass dependence observed in Fig. \ref{fig:loss_scaling}.

\subsection{Adding electric fields and microwaves \label{sec:field}}

The applications of ultracold molecules rely on their permanent dipole moment. 
These dipole moments, however, are not oriented in the lab frame.
To access the molecules' dipole moments, one can apply a static electric field to orient the molecules and thereby align the dipole moments along the field axis \cite{karman:22ref,ospelkaus:10,guo:18}.
Alternatively, microwaves can be used to induce an oscillating dipole moment along the polarization direction,
leading to an equivalent time-averaged interaction \cite{karman:22ref,yan:20}.
Ideally, these molecules that exhibit long-range dipole-dipole interactions would also be stable with respect to collisions.
Above, we have demonstrated that, in the absence of such fields, collisional loss of ultracold molecules can be suppressed by preparing molecules in $j+j'=0+2$ rotational states.
Here, we investigate whether the suppression of loss rates of NaK in $j+j'=0+2$ persists in the presence of external fields. We note that in external fields $j$ is strictly not a good quantum number, and we label the field-dressed states with the value of $j$ approached at zero field.

\begin{figure}
    \centering
    \includegraphics[width=\linewidth]{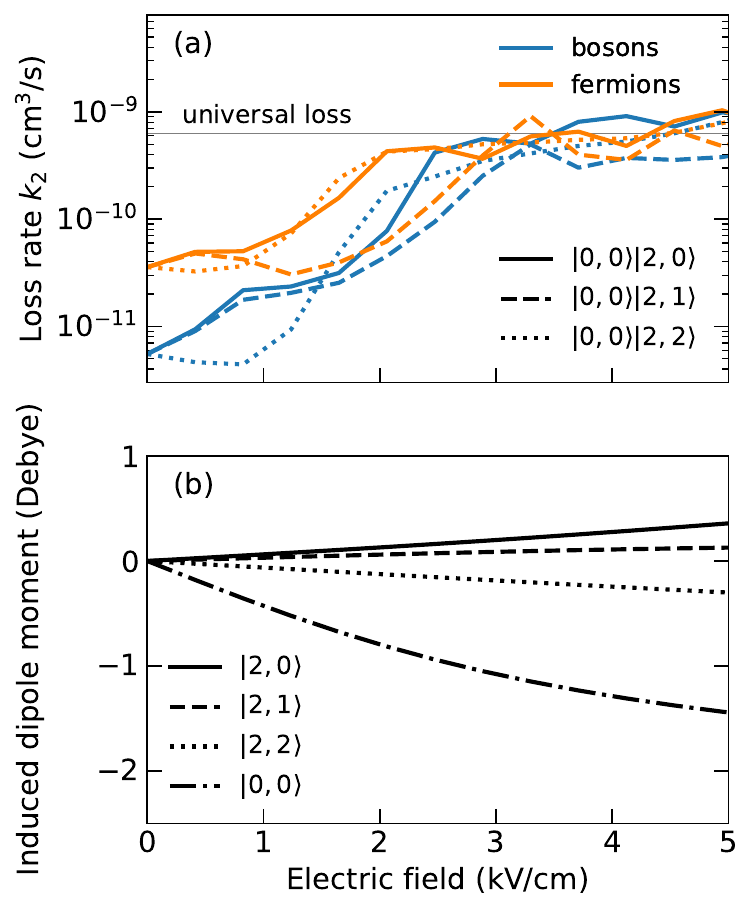}
    \caption{\textbf{The influence of a static electric field} for bosonic and fermionic NaK with $j+j'=0+2$ (denoted at zero field) on (a) the loss rate and (b) the induced dipole moment. The universal loss is given for bosonic NaK.}
    \label{fig:lossE}
\end{figure}

Including a static electric field, we show in Fig.~\ref{fig:lossE}(a) loss rates for bosonic and fermionic NaK $0+2$ states as a function of electric field strength. We find that for field strengths up to around 2 kV/cm both bosons and fermions retain low collision rates until losses rise up to around universal loss again. In Fig.~\ref{fig:lossE}(b) we show the dipole moment that is induced for different rotational states as the electric field strength is increased.
Within the range of low collision rates a fraction of the permanent dipole moment can still be induced, especially for the $j=0$ state.

\begin{figure}
    \centering
    \includegraphics[width=\linewidth]{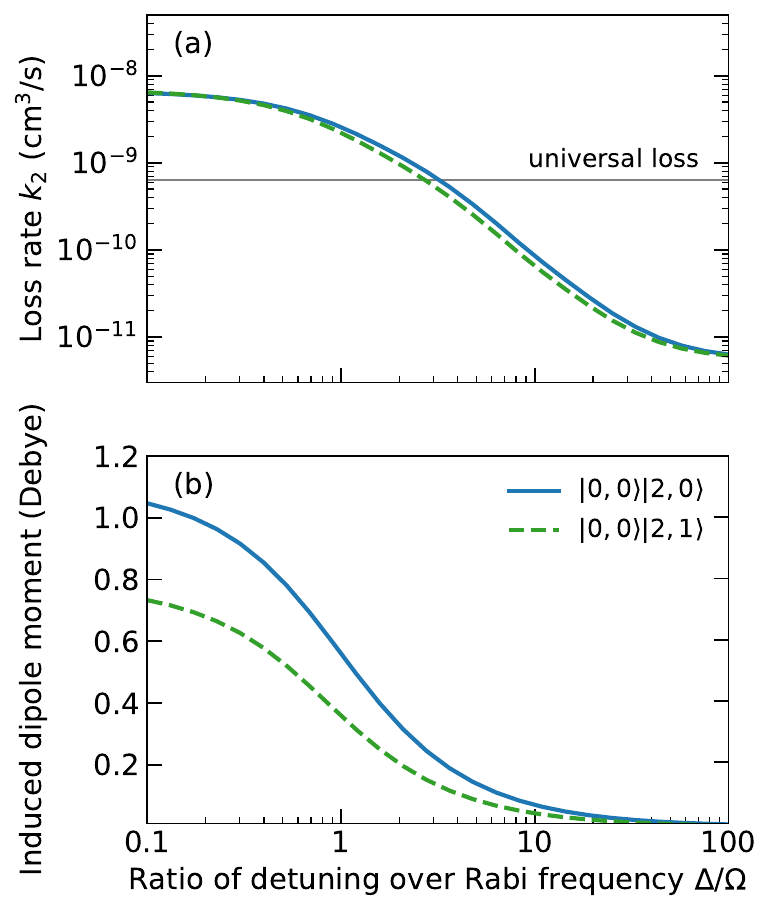}
    \caption{\textbf{The influence of linearly polarized microwaves} for bosonic NaK with rotational $j+j'=0+2$ states (denoted at zero field) on (a) the loss rate and (b) the induced dipole moment, both as a function of microwave detuning $\Delta$ at Rabi frequency $\Omega=2\pi\times10$ kHz.}
    \label{fig:lossMW}
\end{figure}

In Fig.~\ref{fig:lossMW}, we show loss rates for bosonic NaK $0+2$ dressed with microwaves. We use linearly polarized microwaves red detuned from the $j=1\to2$ transition for both $|2,0\rangle$ and $|2,1\rangle$ rotational states \cite{karman:22ref}. We fix the Rabi frequency $\Omega$ to $2\pi\times10$ kHz and scan the detuning $\Delta$ from the transition. The induced dipole moment is then defined as the effective dipole arising from the dipole-dipole interaction \cite{karman:22ref}. The largest dipole moments that can be induced for $|0,0\rangle|2,0\rangle$ and $|0,0\rangle|2,1\rangle$, respectively, are $d/\sqrt{3}$ and $d/\sqrt{6}$. For red-detuned dressing, larger $\Delta/\Omega$ decreases the induced dipole moment by 
\begin{equation}
    d_\mathrm{ind}\sim\frac{1}{\sqrt{1+\left(\frac{\Delta}{\Omega}+\sqrt{1+\left(\frac{\Delta}{\Omega}\right)^2}\right)^2}}\,.
\end{equation}
From Fig.~\ref{fig:lossMW} we then find that relatively small loss rates of $10^{-11}$ to $10^{-10}$ cm$^3$/s can be obtained, while inducing a dipole moment of around 0.1 Debye. Although not shown here, circularly polarized microwaves yield very similar results.

Static electric and microwave fields can still be used to induce a fraction of the dipole moment, while collisional loss is kept low.

\subsection{Interactions at pair degeneracies \label{sec:degenerate_calcs}}

\begin{figure*}
    \centering
    \includegraphics[width=0.9\linewidth]{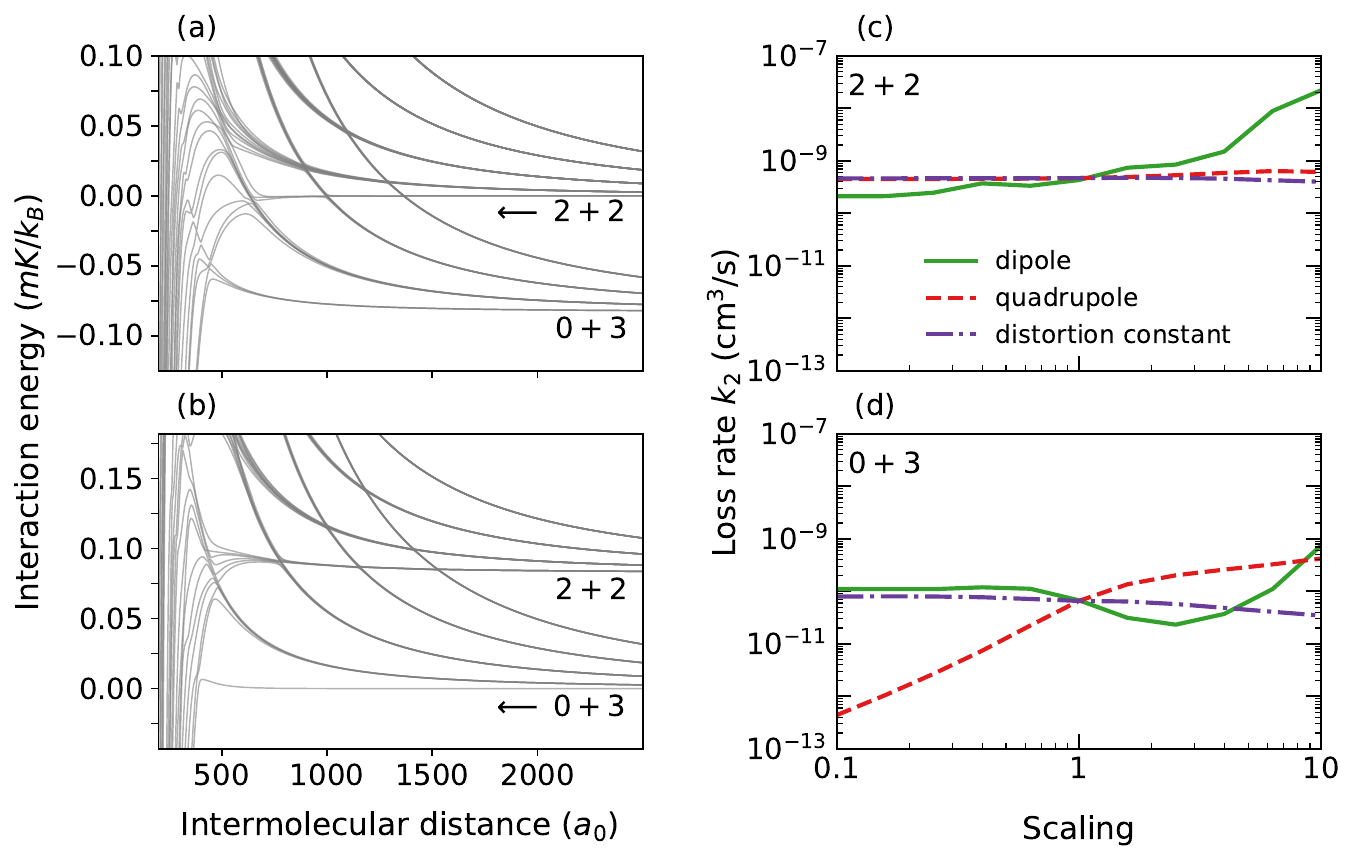}
    \caption{\textbf{Rotational dimer degeneracy} $j+j'=2+2\equiv0+3$ with $m_j=m_{j'}=0$ for bosonic NaK. (a,b) Interaction potentials with centrifugal distortion coefficient taken into account, lifting the degeneracy by $\sim80$ $\mu$K, referenced to the prepared state at infinite separation, which are (a) $2+2$ and (b) $0+3$. The main difference is the parity $(-1)^{j+j'+L}$, which is opposite between these figures. (c,d) Scaling of the loss rates with respect to the dipole moment, quadrupole moment and centrifugal distortion constant for the (c) $2+2$ and (d) $0+3$ states.}
    \label{fig:22}
\end{figure*}

We have discussed in section \ref{sec:degeneracy} that pairs of rotational states contain degeneracies for certain combinations of monomer rotational states. 
The $2+2 \equiv 0+3$ degeneracy is the one with the lowest rotational quantum numbers and we now explore this specific case further, without external fields.
These two pairs of states are coupled by the dipole-quadrupole interaction.
In the absence of that interaction, the $2+2$ state has an attractive rotational van der Waals interaction whereas the $0+3$ state has a repulsive one.
What makes this degeneracy particularly interesting is that the $2+2$ state can involve two molecules prepared in the same $|j=2,m\rangle$ quantum state.
Furthermore, the $0+3$ level is degenerate with $2+2$ for rigid rotors,
but in reality $0+3$ is slightly lower in energy due to centrifugal distortion.
Thus, one can expect level repulsion due to the dipole-quadrupole coupling between the nearly degenerate states to create a repulsive potential correlating adiabatically to $2+2$.
That is, dipole-quadrupole coupling to the nearly degenerate $0+3$ threshold may cause molecules in $j=2$ to be collisionally stable without the need for active shielding.

The interaction curves for $2+2$ and $0+3$ are presented in Figs.~\ref{fig:22}(a) and \ref{fig:22}(b), respectively, where $m_j=m_{j'}=0$. We see the small lifting of degeneracy by about 80 $\mu$K between the two states due to centrifugal distortion of the molecules.
The $L=0$ initial channels for the states $j+j'\equiv2+2$ and $3+0$ have opposite parity,
and the two sets of potentials are shown in panels (a) and (b), respectively.
The rotational van der Waals interaction is attractive in the upper threshold and repulsive in the lower one,
leading to crossings between the potential curves.
The crossings are avoided due to the dipole-quadrupole coupling between the states.
However, the crossings occur where the $2+2$ potential is already attractive and the local kinetic energy makes it unlikely that these crossings would be traversed adiabatically.
For the $0+3$ state, the effect of the avoided crossings is to greatly reduce the barrier height,
which can negatively impact the collisional loss rates.

We compute loss rate coefficients as a function of an artificial scaling applied to the dipole and quadrupole moments as well as the distortion constant and calculate collisional loss for the $2+2$ and $0+3$ initial states, shown in Figs.~\ref{fig:22}(c) and \ref{fig:22}(d), respectively. 
For the $2+2$ state we find high losses across the entire scaling range of the different parameters.
This means that the intrinsic shielding that one could have expected for the $2+2$ state cannot be effective.
The main feature in Fig.~\ref{fig:22}(c) is the rise in loss as the dipole rises, which is simply due to the rotational van der Waals potential becoming more attractive for $2+2$.
For the $0+3$ state rate coefficients are shown in Fig.~\ref{fig:22}(d), and here we find different effects.
For reduced quadrupole moments, the coupling to the $2+2$ state is reduced,
and the loss rate approaches the lower loss rate expected for the $0+3$ repulsive van der Waals potential.
At increased distortion constants, the loss rate decreases slightly as the barrier for $0+3$ increases due to the avoided crossing with the $2+2$ potential being moved outwards to where that potential is less attractive.
Varying the dipole moment changes the repulsive interaction of $0+3$, the attractive interaction of $2+2$ and the coupling between them. The dependence of the loss rate is therefore not as simple.
For all realistic parameters, coupling between the nearly degenerate states is not strong enough to shield the upper state, but it does increase the loss rate for the lower one.

\section{Applications for cold molecules \label{sec:applications}}

The biggest limitation of the stability of ultracold molecular systems is collisional loss.
In ``shielding'', external fields are used to reduce collisional losses \cite{quemener:16,gonzalez:17,matsuda:20,li:21,karman:18d,karman:19c,karman:20,anderegg:21}. 
We discussed in this paper that by an appropriate choice of two rotational states $j,j'$ such that $\Delta j\geq2$ repulsive interactions can be realized without requiring external field control.
Here, we consider several possible applications of these interactions: evaporating a fermi mixture, fermionic molecules in an optical lattice, and stable impurities in an ultracold gas.

\subsection{Evaporation of fermi mixtures \label{sec:evaporation}}

Cooling down to a degenerate quantum gas of fermions has so far been achieved by assembly from atoms at ultracold temperatures \cite{demarco:19,tobias:20,duda:23}, and by forced evaporation in two-dimensional confinement \cite{valtolina:20} or while shielding using static electric fields \cite{matsuda:20,li:21} or microwave fields \cite{anderegg:21,schindewolf:22}.
Efficient forced evaporation requires a high ratio of elastic to inelastic collision rates.
For fermions in the same quantum state, collisions are highly suppressed by the presence of a $p$-wave centrifugal barrier.
This leads to elastic and inelastic collision rates that are suppressed at ultracold temperatures by $T^2$ and $T$, respectively.

Our idea here is that one could create a fermi mixture of molecules in two rotational states, $j$ and $j'$.
Collisions between molecules in the same rotational state would be highly suppressed at ultracold temperatures,
but ``inter-species'' collisions could occur.
If $\Delta j \ge 2$, the inelastic rate could be small due to the repulsive rotational van der Waals interaction.
This would leave only inter-species elastic collisions, which could then enable fast thermalization and efficient forced evaporation.
After removal of molecules in one of the $j$ states, which could be implemented as collisional removal of the minority species, one could end up with a state-pure quantum degenerate gas without active shielding using external fields.

\begin{figure}
    \centering
    \includegraphics[width=\linewidth]{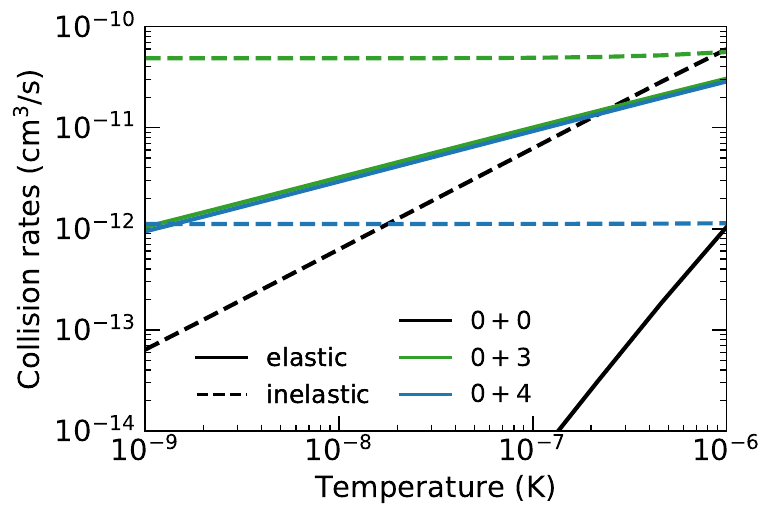}
    \caption{\textbf{Collision rates for thermalization of fermions} for various rotational state pairs of NaK as a function of temperature.}
    \label{fig:thermal}
\end{figure}

In Fig.~\ref{fig:thermal} we show elastic and inelastic collision rates for different pairs of rotational states of fermionic NaK.
The $0+4$ pair performs the best out of the investigated pairs due to the low collisional loss as was seen in Fig. \ref{fig:loss}.
Compared to the $0+0$ collisions, the rate for elastic $0+3$ or $0+4$ collisions has indeed increased by orders of magnitude, while especially for the $0+4$ pair the inelastic rate is generally smaller than the elastic one.
However, the increased elastic rate for $0+4$ is generally not much higher than the inelastic rate of $0+0$ and $4+4$ collisions that also occur in the mixture. 
So while the repulsive interaction improves the elastic collision rate by orders of magnitude while keeping the collisional loss rate low,
the ratio of elastic to inelastic collisions is realistically not high enough to enable direct evaporative cooling.

\subsection{Fermions in optical lattices \label{sec:lattice}}

In quantum simulation using ultracold molecules in optical lattices \cite{yan:13,blackmore:18ref,altman:21},
two molecules can tunnel into the same lattice site and undergo collisional loss.
The repulsive rotational van der Waals interaction can prevent such losses without actively controlling interactions with external fields, at least for fermionic molecules in two rotational states with $\Delta j \ge 2$.
Fermions in the same quantum state will not occupy the same lattice site due to the Pauli principle.
Fermions in distinct quantum states can meet on site, but will be protected from collisional loss by the repulsive rotational van der Waals interaction.
As shown in Sec.~\ref{sec:field}, a weak microwave or static electric field can be applied to access dipolar interactions between the molecules without compromising the collisional stability.

\subsection{Impurity physics \label{sec:impurity}}

An ultracold mixture of molecules in different rotational states is also an interesting playground for impurity physics \cite{massignan:14,perez:21,schmidt:15},
and enables studying polarons with long-range interactions.
We envision introducing $j=2$ impurities into a gas of $j=0$ rotational ground state molecules,
such that the repulsive rotational van der Waals interaction would render the impurities stable with respect to collisions with the bath.
One can address the impurity by microwaves tuned on the $j=1\to2$ transition,
which induces dipole-dipole interactions between the impurity and the bath molecules selectively,
without inducing dipolar interactions within the bath.

\section{Conclusion \label{sec:conclusion}}

In this work, we described the electrostatic interaction between two dipolar, linear molecules that are prepared in specific rotational states.
When both molecules are prepared in the same rotational state, we recover the well-known attractive rotational van der Waals interaction.
When the rotational states differ by one quantum, resonant dipole-dipole interactions lead to first-order dipole-dipole interactions.
When the rotational states differ by two or more quanta, the dominant interaction is a repulsive van der Waals interaction.
The repulsive interactions follow directly from the rotational energy levels of a linear molecule that become sparser as one goes up in energy,
and are therefore universal for polar molecules.
This is in stark contrast to the usual electronic van der Waals interaction which is typically attractive as generally electronically excited states become denser at higher energies.

The repulsive interactions result in a potential barrier that we computed for NaK and KCs molecules.
These barriers can suppress collisional losses at low temperatures.
Using quantum scattering calculations we studied the ultracold collision dynamics quantitatively and showed that collisional loss can be suppressed by several orders of magnitude below the universal loss rate.
Moreover, we discussed the cases when pairs of rotational energy levels of two molecules become degenerate, which impacts the collision dynamics.
Using static electric fields or microwaves, a fraction of the limiting dipole moment can be induced while maintaining a low loss rate.

The repulsive interaction between molecules in different rotational states opens up several applications in ultracold physics.
A clear application is quantum simulation with fermionic molecules in an optical lattice.
Here, only distinguishable fermionic molecules in different rotational states can tunnel into a single lattice site,
and the repulsive van der Waals interactions can prevent collisional loss and control on site interactions.
Another promising application is impurity physics with rotationally excited molecules immersed in a bath of rotational ground state molecules,
enabling control of dipolar and repulsive van der Waals interactions between the impurity and bath molecules while suppressing collisional loss.
Further less obvious applications of the repulsive rotational van der Waals interactions may exist.
In the accompanying paper we show how these interactions can be used to implement near-deterministic loading of optical tweezers with laser cooled molecules \cite{walraven:24b}.

\end{document}